\documentclass[letter]{aa} % for the letters 
\usepackage{graphicx}
\usepackage[utf8]{inputenc}
\usepackage[english]{babel}
\usepackage{lscape}
\usepackage{caption}
\usepackage{longtable}
\usepackage{xcolor}
\usepackage{rotating}
\usepackage{float}
\usepackage{subfig}
\usepackage{multicol} 
\usepackage{txfonts}
\usepackage{hyperref}
\begin{document}

   \title{Gas, not dust: Migration of TESS/\textit{Gaia} hot Jupiters possibly halted by the magnetospheres of protoplanetary disks} 
   \titlerunning{Migration of hot Jupiters halted by the magnetospheres of protoplanetary disks}

   %\subtitle{}

   \author{I. Mendigutía\inst{1}
   \and
   J. Lillo-Box\inst{1}
   \and
   M. Vioque\inst{2,3}
   \and
   J. Maldonado\inst{4}
   \and
   B. Montesinos\inst{1}
   \and
   N. Huélamo\inst{1}
   \and
   J. Wang\inst{5}
   %\and
   %E. Villaver\inst{4}
   %et al.
   }

   \institute{$^{1}$Centro de Astrobiología (CAB), CSIC-INTA, Camino Bajo del Castillo s/n, 28692, Villanueva de la Cañada, Madrid, Spain.\\
   $^{2}$European Southern Observatory, Karl-Schwarzschild-Strasse 2, D-85748 Garching bei München, Germany\\
   $^{3}$Joint ALMA Observatory, Alonso de Córdova 3107, Vitacura, Santiago 763-0355, Chile\\
   $^{4}$INAF – Osservatorio Astronomico di Palermo, Piazza del Parlamento 1, I-90134 Palermo, Italy\\
   $^{5}$Departamento de F\'isica Te\'{o}rica, M\'{o}dulo 15, Facultad de Ciencias, Universidad Aut\'{o}noma de Madrid, 28049 Madrid, Spain
   }

   \date{Received January 28, 2024; accepted April 27, 2024}

% \abstract{}{}{}{}{} 
% 5 {} token are mandatory
 
  \abstract
   {The presence of short-period (< 10 days) planets around main sequence (MS) stars has been associated either with the dust-destruction region or with the magnetospheric gas-truncation radius in the protoplanetary disks that surround them during the pre-MS phase. However, previous analyses have only considered low-mass FGK stars, making it difficult to disentangle the two scenarios. }
   {This exploratory study is aimed at testing whether it is the inner dust or gas disk driving the location of short-period, giant planets.}
   {By combining TESS and \textit{Gaia} DR3 data, we identified a sample of 47 intermediate-mass (1.5 -- 3 M$_{\odot}$) MS stars hosting confirmed and firm candidate hot Jupiters. We compared their orbits with the rough position of the inner dust and gas disks, which are well separated around their Herbig stars precursors. We also made a comparison with the orbits of confirmed hot Jupiters around a similarly extracted TESS/\textit{Gaia} sample of low-mass sources (0.5 -- 1.5 M$_{\odot}$).}
   {The orbits of hot Jupiters around intermediate-mass stars tend to be closer to the central sources than the inner dust disk, most generally consistent with the small magnetospheric truncation radii typical of Herbig stars ($\lesssim$ 5R$_*$). A similar study considering the low-mass stars alone has been less conclusive due to the similar spatial scales of their inner dust and gas disks ($\gtrsim$ 5R$_*$). However, considering the whole sample, we do not find the correlation between orbit sizes and stellar luminosities that is otherwise expected if the dust-destruction radius limits the hot Jupiters' orbits. On the contrary, the comparative analysis reveals that such orbits tend to be closer to the stellar surface for intermediate-mass stars than for low-mass stars, with both being mostly consistent with the rough sizes of the corresponding magnetospheres.}
   {Our results suggest that the inner gas (and not the dust) disk limits the innermost orbits of hot Jupiters around intermediate-mass stars. These findings also provide tentative support to previous works that have claimed this is indeed the case for low-mass sources. We propose that hot Jupiters could be explained via a combination of the core-accretion paradigm and migration up to the gas-truncation radius, which may be responsible for halting inward migration regardless of the stellar mass regime. Larger samples of intermediate-mass stars with hot Jupiters are necessary to confirm our hypothesis, which implies that massive Herbig stars without magnetospheres ($>$ 3-4 M$_{\odot}$) may be the most efficient in swallowing their newborn planets. }


   \maketitle
%
%-------------------------------------------------------------------

\section{Introduction}
\label{Sect:intro}
Short-period (< 10 days) planets orbiting close (< 0.1 au) to their host, main sequence (MS) stars are relatively uncommon \citep{Kunimoto20}. However, their role in shaping our understanding of planet formation has been recognized since the first detection of 51 Peg b \citep{Mayor95}, especially with respect to the gas giants known as "hot Jupiters." Without neglecting the role that interactions with other stars and planets may play in explaining hot Jupiters around some systems, in this work we assume that their orbits are primarily fixed by the physical conditions of the protoplanetary disks that surround young stars during the pre-MS \citep[see below and e.g.,][]{Mulders15,Benkendorff24}.

Protoplanetary disks are limited by two main barriers that prevent them from extending up to the stellar surfaces \citep[e.g.,][]{Romanova19}. Solid particles cannot survive close to the stars because dust sublimates at temperatures above $\sim$ 1500 K. The size of such a dust barrier scales with the square root of the stellar luminosity \citep{Tuthill01,Monnier02}. Gas in disks is also truncated close to the central sources, where it is channeled by the star's magnetic field until it accretes at high stellar latitudes. The size of this magnetospheric barrier for the gas decreases for smaller stellar magnetic fields and stronger mass accretion rates \citep{Koenigl91}. 

Regardless of whether planets form in situ or migrate, the limits established by both the inner dust and gas radii have been invoked to explain the presence of short-period planets and its abrupt decline at smaller orbits \citep[e.g.,][]{Lee17,Flock19,Romanova19}. In particular, the role that the  magnetospheric truncation radius may play as a nearly universal explanation for close orbits in widely different systems was  recently pointed out \citep{Batygin23}. However, previous analyses focused on low-mass FGK stars, making hard to observationally disentangle which of the two barriers, dust or gas, actually determines the innermost planetary orbits. Indeed, the magnetospheric truncation radii in solar type and lower-mass, T Tauri stars are typically similar to their dust-destruction radii \citep[5-10 R$_*$ or $\sim$ 0.05 au;][]{Pinte08,Bouvier07,Wojtczak23}, both associated with comparable orbital periods. 

\begin{figure}
   \centering
   \includegraphics[width=9cm]{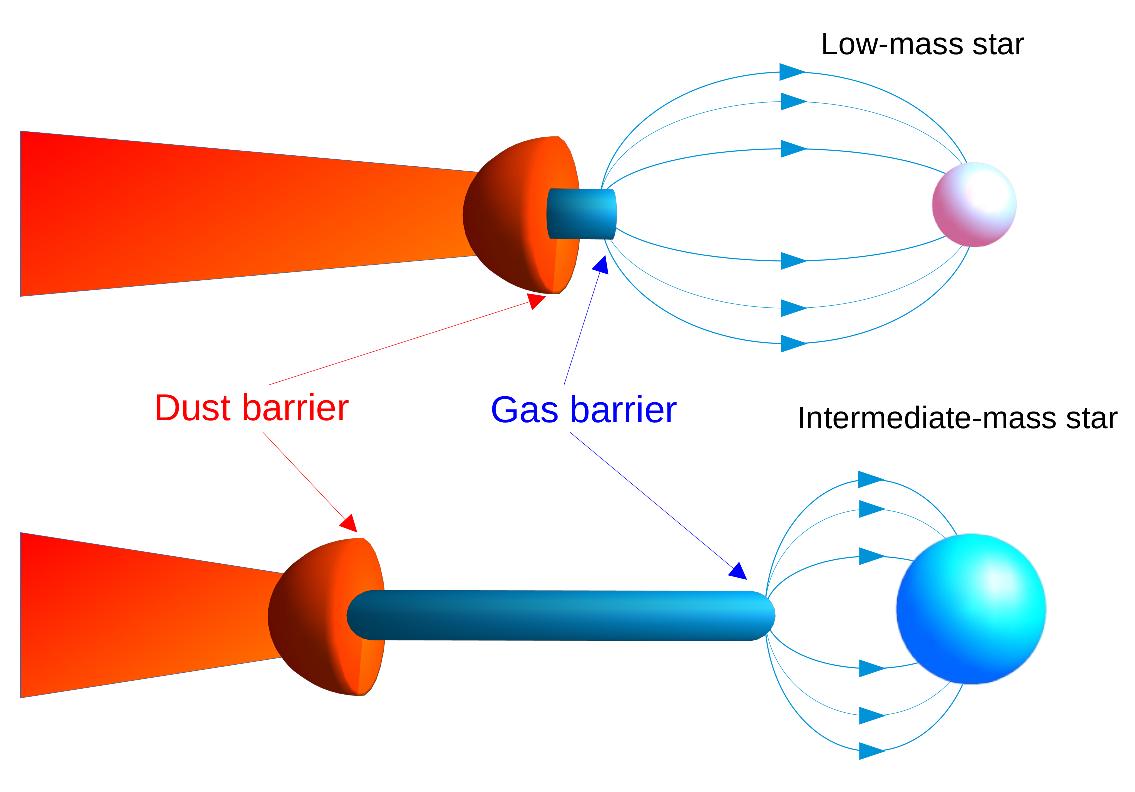}
      \caption{Sketch showing an edge-on dust (red) and gas (blue, arrows indicate the magnetospheres) protoplanetary disk around low-mass (top) and intermediate-mass (bottom) young stars. The separation between the inner dust and gas disks is typically larger in intermediate-mass stars because of their higher luminosities and weaker magnetic fields. \citep[Adapted from][]{Mendi20}.} 
         \label{fig:sketch_dust_gas}
   \end{figure}

On the other hand, from the study of stellar interiors it is well settled that convective envelopes necessary to sustain strong, ordered magnetic fields are small or absent in MS stars with spectral type A and earlier \citep[e.g.][and references therein]{Simon02}. For this reason (and because mass accretion rates increase with the stellar mass), the magnetospheric radius tends to significantly reduce or even disappear (< 5R$_*$) in their younger precursors, the Herbig stars \citep[see, e.g., the reviews in][and references therein]{Mendi20,Brittain23}. In addition, the inner dust disk sizes of Herbig stars are well constrained from interferometric observations, being at least one order of magnitude larger than in T Tauri stars \citep[e.g.,][]{MarcosArenal21}. Because the difference between the spatial location of the inner gas and dust disks is generally much larger in Herbig stars than in T Tauri stars, the best way to address which one fixes the minimum planetary orbits is by including in the analysis short-period planets around MS stars with intermediate masses. In this work, we follow the classical boundary of 1.5 M$_{\odot}$ to divide between low- and intermediate-mass MS stars with and without convective sub-photospheric regions (however, also see Sect. \ref{Sect:discussion}), for which  statistical differences at a population level are expected. In particular, if the dust barrier controls the innermost planetary orbits, then the minimum star-planet separations should scale with the square root of the stellar luminosity and, thus, be typically larger for intermediate-mass stars than for less luminous, low-mass stars. On the contrary, if the gas barrier dominates then intermediate-mass stars should host planets with orbits generally closer to the stellar surfaces than those around lower mass stars \citep[see Fig. \ref{fig:sketch_dust_gas} and e.g.,][for a supplementary discussion]{Lee17,Batygin23}. 

Nevertheless, the number of confirmed, short-period planets around intermediate-mass MS stars is still very scarce (Sect. \ref{Sect:sample}). A major issue is that the confirmation of planetary candidates, commonly identified through the transit method, usually requires additional detections through radial velocity analyses based on Doppler shifts. However, intermediate-mass stars show a limited amount of (rotationally broadened) absorption lines, for which radial velocity analyses are generally not applicable. Alternatively, statistical validation algorithms make use of light curves and ancillary observations to rule out false-positives \citep[e.g.,][and references therein]{Giacalone21}. Because of the large photometric apertures used by the Kepler mission \citep{Borucki10}, and especially by the Transiting Exoplanet Survey Satellite \citep[TESS,][]{Ricker15}, the key task of the previous validation algorithms is to discard the presence of eclipsing binaries in the field that may mimic a planetary transit around the target star. For this purpose, the binarity information that \textit{Gaia} provides for hundreds of thousands of sources \citep{Prusti16,Vallenari23,Arenou23} constitutes an alternative way to detect false positives \citep[e.g.,][]{Tarrants23}. Moreover, \textit{Gaia} DR3 provides stellar characterization for more than a billion sources \citep{Creevey23,Fouesneau23}, which constitutes a great tool to homogeneously analyze populations.   

We follow the latter approach to study the influence of the protoplanetary disk barriers on the location of short-period planets. TESS has targeted stars brighter than those observed by Kepler, covering a significant number of A stars with a cadence and sensitivity that is ideal for detecting short-period, giant planets \citep[e.g.,][]{Zhou19}. In this work, we combine TESS and \textit{Gaia} DR3 data to analyze a sample of hot Jupiters around MS, intermediate-mass stars. Section \ref{Sect:sample} describes the sample selection and its properties. Section \ref{Sect:results} explores the trends shown by short-period planets around intermediate-mass stars, compared with the rough location of the dust and gas barriers and with an analogous sample of hot Jupiters around low-mass stars. Section \ref{Sect:discussion} discusses the previous results and Sect. \ref{Sect:conclusions} offers a summary of our conclusions.  

\section{Sample selection and properties}
\label{Sect:sample}
This work is based on sources with exoplanet candidates identified from TESS light curves or "TESS objects of interest" \citep[TOIs;][]{Guerrero21}. The list of TOIs was filtered and cross-matched with \textit{Gaia} DR3 data to select the sample. Details of the selection process are included in Appendix \ref{appendix_sample_selection}, but we offer a brief summary below. 

Essentially, we selected the most reliable TOIs with short-period planets (P $<$ 10 days) and \textit{Gaia} DR3 stellar characterization, including the evolutionary state, luminosity, mass, and radius (L$_*$, M$_*$, R$_*$). The \textit{Gaia} parameters allowed us to disentangle the MS, intermediate-mass host stars from the rest. A careful filtering process was carried out to prevent contamination of the light curves by additional stars within the TESS apertures. Although the TOIs rejected could include planets that may be validated or confirmed in the future (e.g., Lillo-Box et al. in prep), the strict filtering criteria applied lead to a robust sample of 47 intermediate-mass, MS stars. In particular,  25 such stars host confirmed planets and 22 have planet candidates with minimal chances of turning out to be false positives.

Appendix \ref{appendix_properties_sample} summarizes the properties of this sample. The stellar masses range from 1.5 to 3 M$_{\odot}$ and the planetary sizes extend from Neptune to a few Jupiter radii. The parameter space covered by the 25 confirmed and 22 candidate planets is similar and the general results from this work do not change when considering each sub-sample separately. All 47 sources are studied to increase the statistical significance and the term "planet" or "candidate planet" is used interchangeably in the rest of the manuscript.  

We also consider 298 low-mass, MS stars (0.5 $<$ M$_*$/M$_{\odot}$ $\leq$ 1.5) included both in the TOIs and \textit{Gaia} DR3 catalogs. All of them host confirmed, short-period giant planets similar to those around the intermediate-mass sample. This control sample of low-mass stars is used mainly for comparison purposes, maintaining the homogeneity of the analysis based on TESS and \textit{Gaia} data. Although other samples of low-mass stars with hot Jupiters have been studied in the past, the distributions of orbital periods and radii presented here are similar to previous works. Details on the low-mass sample selection and its properties are also given in Appendices \ref{appendix_sample_selection} and \ref{appendix_properties_sample}. In addition, Appendix \ref{appendix_properties_sample} provides support to the underlying assumption that potential observational biases similarly affect the intermediate- and low-mass samples, for the range of orbital periods and planet sizes analyzed. Thus, such biases are not expected to drive the trends presented in this work.

\section{Results}
\label{Sect:results}

We are interested on the relative positions of the planets with respect to the protoplanetary disk dust and gas barriers, which are defined as a function of pre-MS stellar luminosities and radii. Appendix \ref{appendix_pre-MS} details how the pre-MS values for L$_*$ and R$_*$ have been inferred for the stars in the sample. In short, the luminosities and radii exhibited by the stars throughout the pre-MS were derived from the crossing points between the evolutionary tracks for each stellar mass and the 3 Myr theoretical isochrone in \citet{Siess00}. Such a young age is the typical timescale when most protoplanetary disks dissipate \citep[e.g.,][and references therein]{Mendi22}. As discussed in Appendix \ref{appendix_pre-MS}, our general results and conclusions are not significantly affected by the use of different disk dissipation timescales, theoretical evolutionary tracks, or isochrones.

\begin{figure}
   \centering
   \includegraphics[width=9cm]{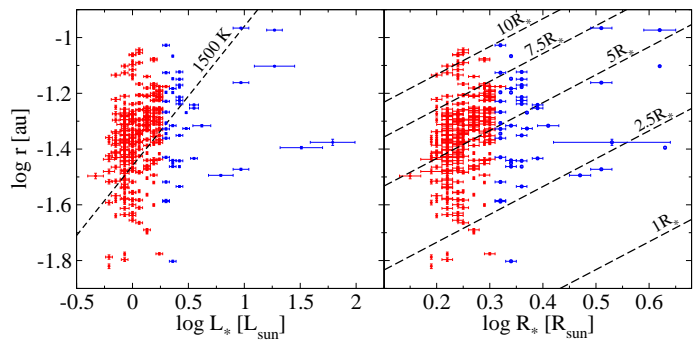}
      \caption{Planetary orbital radii versus pre-MS stellar luminosities (left) and radii (right) at 3 Myr. Intermediate- and low-mass stars are displayed in blue and red, respectively. In the left panel, the dashed line indicates the inner dust disk for a dust sublimation temperature of 1500 K. In the right panel, the dashed lines indicate the magnetospheric inner gas disk at 10, 7.5, 5, 2.5, and 1R$_*$.}
      \vspace{0.85cm}
         \label{fig:orbradius_gas_dust_preMS}
   \end{figure}

Figure \ref{fig:orbradius_gas_dust_preMS} (left) shows the planetary orbital radii, obtained from the TESS orbital periods, \textit{Gaia} DR3 stellar masses, and Kepler's third law -- versus the pre-MS stellar luminosities. The dashed line is the L$_*^{1/2}$ scaling law limiting the inner dust disk radius as derived by \citet{Monnier02} for a typical dust sublimation temperature of 1500 K. Interferometric observations of inner dust disk sizes for T Tauri and Herbig stars indicate that essentially all them are consistent with the previous limit \citep[][and references therein]{MarcosArenal21}. In contrast, $\sim$ 70$\%$ of the intermediate-mass stars have planets with orbits closer to the central source than the dust barrier.

Figure \ref{fig:orbradius_gas_dust_preMS} (right) shows the planetary orbital radii versus the pre-MS stellar radii. The dashed lines indicate the orbits at 10, 7.5, 5, 2.5, and 1R$_*$. The location of $\sim$ 60$\%$ of the planets around intermediate-mass stars is consistent with the small magnetospheres typical of Herbig stars ($<$ 5R$_*$), being the rest orbiting at larger distances.

Concerning the sample of low-mass stars, Fig. \ref{fig:orbradius_gas_dust_preMS} shows that almost 80$\%$ of such sources host planets in orbits equal or larger than the dust-destruction radius, and $\sim$ 65$\%$ have orbits $>$ 5R$_*$. This illustrates the notion noted in Sect. \ref{Sect:intro}: the dust and gas barriers roughly coincide in low-mass stars, with the majority of their planets located in orbits that are consistent both with the dust-destruction radius and with the large magnetospheric gas-truncation radii typical for T Tauri stars.

\begin{figure}
   \centering
   \includegraphics[width=9cm]{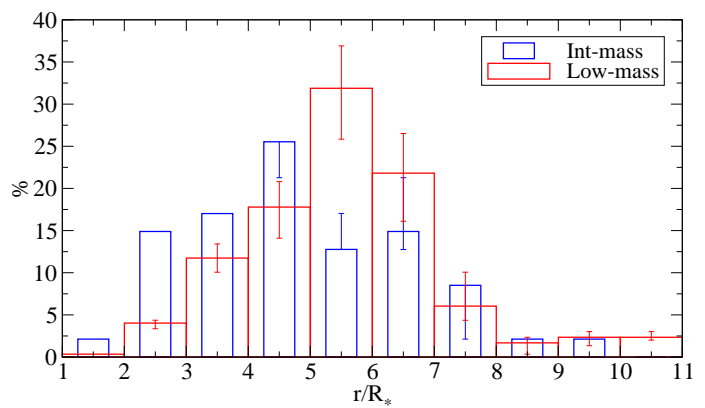}
      \caption{Distributions of planetary orbital radii in terms of of pre-MS stellar radii at 3 Myr. The intermediate- and low-mass samples are in blue and red, as indicated in the legend} 
         \label{fig:histogram_preMS}
   \end{figure}

Figure \ref{fig:histogram_preMS} shows the above mentioned statistical differences between the distributions of planetary orbits, which are expressed as a function of pre-MS stellar radii. The orbits around intermediate- and low-mass stars dominate below and above 5R$_*$, respectively, which is the rough limit dividing between small and large magnetosphere sizes. A two-sample Kolmogorov-Smirnov (K-S) test provides a 0.53$\%$ probability that the previous samples are drawn from the same parent distribution.

\section{Discussion}
\label{Sect:discussion}

The previous statistical results are affected by several uncertainties. Although 1.5M$_{\odot}$ has been used to divide between low- and intermediate-mass MS stars with and without convective sub-photospheric regions, this is not a sharp division \citep[e.g.,][]{Cantiello19}. Moreover, the extent of convective sub-photospheric regions changes during the pre-MS evolution of stars in the mass range $\sim$ 1 -- 3 M$_{\odot}$ \citep[e.g.,][]{Hussain14}. In addition, different calculations of the L$_*^{1/2}$ scaling law for the dust-destruction radius provide different values for the intercept, which mainly depend on the assumed dust sublimation temperature and composition\footnote{However, expressions with different slopes departing from the L$_*^{1/2}$ scaling law mainly refer to stars more massive and luminous than those considered in this work; see e.g. \citet{MarcosArenal21} and references therein.} \citep[e.g.,][]{Koumpia21}. Similarly, as mentioned in the introduction, the exact position of the gas-truncation radius also varies from star to star, for which the 5R$_*$ limit is only a first-order approximation \citep[e.g.,][]{Wojtczak23}. Therefore, the statistics provided in the previous section should be considered merely as a rough description of the behaviour of intermediate- and low-mass populations and their differences. 

With the previous caveats and the limitations of our sample in mind, our results suggest that the small inner gas disks of Herbig stars (and not the inner dust disks) act as effective barriers, effectively determining the innermost planetary orbits for the majority of intermediate-mass stars. All such stars host a giant-type planet with the size of Neptune or larger (Sect. \ref{Sect:sample} and Fig. \ref{fig:planets_rad_period}). Based on the approach that their observed orbits remain relatively stable after disks dissipate, two possible explanatory scenarios are explored. First, the giant planets are fully made of gas, opening up the possibility that they formed in situ within the dense inner gas disks where no solid content is present. Second, the giant planets have a solid core massive enough to accrete gas, requiring their formation to begin at distances no nearer than the dust barrier and then migrated inward. 

The first scenario is in line with planet formation models through gravitational instabilities \citep[e.g.,][]{Boss98}. However, the formation of planets via the fragmentation of self-gravitating protoplanetary disks requires both large masses and low temperatures. The last condition could only be achieved in the optically thin outer regions of the disks \citep[> 50 au;][]{Gammie01}. Thus, although the formation of giant planets through gravitational instability could actually occur in the outer disks of Herbig stars \citep{Dong18}, it is unrealistic to assume that they also form in this way in the hot inner disks. Moreover, the amount of gas available within the first $\sim$ 0.1 au is not enough to form Jupiter-like planets \citep{Mulders21,Zhu23}. On the other hand, that giant planets formed trough gravitational instabilities migrate from tens of au to just a few stellar radii does not seem realistic either. In fact, related models and simulations tend to use computational grids with much larger inner edges \citep[e.g.,][and references therein]{Baruteau11,Michael11}.

The most plausible alternative explanation is that the innermost giant planets around intermediate-mass stars started their formation beyond the dust barrier, where it is possible to build a solid core capable of accreting substantial amounts of gas before and during their migration up to the gas barrier. This formation mechanism is analogous to the classical core-accretion scenario, combined with migration up to a few stellar radii, which has been invoked to explain the presence of hot Jupiters around low-mass stars \citep[e.g.,][]{Lin96,Nelson00,Kley12,Drkazkowska23,Zhu23}. 

Concerning the sample of low-mass stars, we cannot unambiguously disentangle which of the two protoplanetary disk barriers, gas or dust, is most likely to determine the innermost planetary orbits, at least based on that sample alone. Indeed, the comparison with the sizes of the inner dust and gas disks typical of low-mass young stars is not conclusive, as both roughly coincide in this type of source and are similar to most observed planetary orbits. However, the extrapolation of our results in the context of intermediate-mass stars is in line with previous claims suggesting that the magnetospheres truncating the gas disks constitute the limit of innermost planetary orbits also for lower-mass sources \citep{Mulders15,Lee17,Batygin23}. Moreover, although the whole sample spans over orders of magnitude in stellar luminosites, the $\sim$ L$_*^{1/2}$ correlation with the orbital size (expected if the dust-destruction radius was limiting planetary orbits) has not been observed (left panel of Fig. \ref{fig:orbradius_gas_dust_preMS}). In contrast, that the observed orbits tend to be closer to the stellar surfaces for intermediate-mass stars than for low-mass stars, both being mostly consistent with the rough sizes of the corresponding magnetospheres (right panel of Fig. \ref{fig:orbradius_gas_dust_preMS}, and Fig. \ref{fig:histogram_preMS}), may be indicating that these are responsible for ceasing inward migration regardless of the stellar mass regime.

\section{Concluding remarks}
\label{Sect:conclusions}
%Based on direct imaging of long-period planets and radial velocity analyses of post-MS stars, the occurrence rate of giant planets peaks in the intermediate-mass stellar regime and drops drastically for masses $>$ 3M$_{\odot}$ \citep[][]{Reffert15,Wagner2022,Wolthoff22}. Thus, understanding planet formation requires the assessment of intermediate-mass stars. Although recent theoretical efforts on planet formation are devoted to that mass regime \citep{Ronco23,Johnston24}, we still lack from a solid observational background on exoplanets analogous to that around low-mass, MS stars. This abrupt observational gap affects our knowledge on planet formation in general and on hot Jupiters in particular. 
The lack of a solid observational background on exoplanets around intermediate-mass stars (analogous to the background present around low-mass stars) affects our knowledge of planet formation in general and of hot Jupiters in particular. In this work we have analysed such type of planets, selected from a combination of TESS and \textit{Gaia} data. Our analysis has focused on the physical limit of their innermost planetary orbits, regardless of the ongoing debate on the frequency of hot Jupiters around intermediate-mass stars \citep[e.g.,][and references therein]{Sebastian2022}. In principle, our analysis is also independent of observational developments that would eventually lead to larger samples of intermediate-mass stars hosting smaller exoplanets at longer orbital radii. We provide tentative evidence to support the notion that hot Jupiters' orbits around intermediate-mass stars are mostly determined by the protoplanetary disk gas-truncation radius -- and not by the dust-destruction radius. Although gravitational instabilities may play a role in the formation of long-period giant planets around such stars, we have suggested that the origin of hot Jupiters is probably similar than it is for lower-mass sources. This is based on a combination between the core-accretion paradigm and migration up to the inner gas edge. Finally, the comparison between low- and intermediate-mass stars suggests that the gas barrier indeed fixes the innermost planetary orbits for the whole stellar mass regime. Future tests of the previous hypothesis require larger samples of intermediate-mass stars with hot Jupiters. Two examples of such types of tests are outlined below.

First, the size of the magnetosphere is limited by the disk co-rotation radius, which is smaller for larger stellar rotational velocities \citep{Shu94}. Thus, if the magnetosphere controls the innermost planetary orbits these should be smaller for fast-rotating stars \citep[see, e.g., the related discussion in][]{Lee17}. This is in agreement with the recent finding, showing that shorter orbital periods are observed in more massive stars with shorter rotational periods, at least considering FGK spectral types \citep{Garcia23}. However, it is hard to make a conclusive test only based on low-mass stars, given their narrow range of small projected rotational velocities. In contrast, velocities of intermediate-mass stars span from a few to a few hundred km/s, making them ideal for such a test. \textit{Gaia}-based projected rotational velocities are presently only available for a dozen of all the sources analyzed in this work. Additional velocity estimates will be helpful in carrying out this task. 

Second, that magnetospheres act as gas barriers ceasing inward migration immediately implies that if those are absent, then the probability that planets are swallowed by their host stars increases \citep{Nelson00}. Indirect evidence of planets swallowed by their hosts have been provided only for a few solar-type stars \citep[e.g.,][and references therein]{Israelian01,De23}. Notably, magnetospheres are likely to be lacking in most Herbig stars with masses $\gtrsim$ 3-4 M$_{\odot}$ \citep{Wichittanakom20,Vioque22}, for which the gas disk may reach the central source trough a boundary layer \citep[][and references therein]{Mendi20}. Thus, if magnetospheres are the ultimate barrier preventing unlimited planet migration, then the planet engulfment scenario would be most efficient for stellar masses of $>$ 3-4 M$_{\odot}$. These stars may show a deficit of hot Jupiters, as compared to the case of less massive stars.  

\begin{acknowledgements}
The authors acknowledge the anonymous referee, whose suggestions have served to improve the manuscript. IM's research is funded by grants PID2022-138366NA-I00, by the Spanish Ministry of Science and Innovation/State Agency of Research MCIN/AEI/10.13039/501100011033 and by the European Union, and by a  Ram\'on y Cajal fellowship RyC2019-026992-I. J.L.-B. is partly funded by the Spanish MCIN/AEI/10.13039/501100011033 and NextGenerationEU/PRTR grants PID2019-107061GB-C61 and CNS2023-144309, and by the Ram\'on y Cajal fellowship RYC2021-031640-I. BM is supported by grant MCIN/AEI/PID2021-127289-NB-I00. We acknowledge the use of public TOI Release data from pipelines at the TESS Science Office and at the TESS Science Processing Operations Center. Funding for the TESS mission is provided by NASA’s Science Mission directorate.
This work has made use of data from the European Space Agency (ESA) mission
{\it Gaia} (\url{https://www.cosmos.esa.int/gaia}), processed by the {\it Gaia}
Data Processing and Analysis Consortium (DPAC,
\url{https://www.cosmos.esa.int/web/gaia/dpac/consortium}). Funding for the DPAC
has been provided by national institutions, in particular the institutions
participating in the {\it Gaia} Multilateral Agreement.
\end{acknowledgements}

\bibliographystyle{aa}
\bibliography{myrefs.bib}

\newpage
\begin{appendix} 
%\onecolumn
\section{Sample selection}
\label{appendix_sample_selection}
The description of the TOIs catalog is in \citet{Guerrero21}, but the updated list including 6977 TOIs\footnote{https://tev.mit.edu/data/} was taken as a departure point for selecting the sample. The previous list was initially filtered by only considering the sources that according with the TOIs catalog have: i) A light curve that is compatible with planetary transits (i.e., not rejected as false positives, not related with instrumental noise, or not classified as eclipsing binaries, with sources showing centroid offsets also discarded, stellar variables, or ambiguous planet candidates). ii) An orbital period of P < 10 days. For the few exceptions where two or more planets are associated to a given star, only the one with the shortest period was considered.

The resulting list was cross-matched with the \textit{Gaia} DR3 catalog of astrophysical parameters produced by the Apsis processing \citep{Creevey23,Fouesneau23}. A radius of 1" was used to cross-match the TOIs and \textit{Gaia} coordinates. A second filter was then applied, keeping the single sources that have: iii) \textit{Gaia} DR3 values for L$_*$, M$_*$, and R$_*$. iv) An evolutionary status consistent with being in the MS (i.e., \textit{Gaia} DR3 "evolution stage" parameter of $\le$ 420). After the previous process, the initial number of TOIs is reduced by more than a factor of 3. Among the resulting list, 25 intermediate-mass stars with light curves having either a "confirmed planet" (CP) or "Kepler planet" (KP) status in the TOIs catalog were identified. These are all TOIs with \textit{Gaia} DR3 masses > 1.5M$_{\odot}$ and confirmed short-period planets (i.e., less massive than 13 M$_{Jup}$) currently identified in the Encyclopaedia of Exoplanetary Systems\footnote{https://exoplanet.eu/home/}. This sample was kept and the remaining intermediate-mass stars were further filtered as follows. 

To reject false positives due to additional eclipsing binaries, the previous list was cross-matched with the \textit{Gaia} DR3 catalog of Non-Single Stars \citep[NSS;][]{Arenou23}. Although this distinguishes between eclipsing binaries and other types of multiple systems, the cross-match was done with the entire \textit{Gaia} NSS catalog. This way, the possibility that the TESS apertures are contaminated by unrevealed eclipses (mainly in spectroscopic binaries) is prevented. A cross-match radius of 63" was used, which corresponds to a TESS aperture of $\sim$ 6 $\times$ 6 pixels. Such an aperture is similar to the ones typically used to extract the light curves of the TOIs from the SPOC and QLP pipelines \citep[][and references therein]{Guerrero21}. Around a hundred of actual and potential eclipsing binaries coincident or close to the TOIs were removed through this process. Finally, the same radius was used to cross-match the remaining TOIs with the \textit{Gaia} DR3 Main Source catalog \citep{Vallenari23}. This lists the \textit{Gaia} Rp magnitudes, with passbands that are most similar to TESS'. All TOIs with flux contamination $\sum$ F$_i$/F$_*$ > 5$\%$ from stars in the aperture were then eliminated. This last filter is the most stringent, removing several hundreds TOIs. Although the general results of this work do not change by using or not the previous filter, we applied it to minimize the possibility that new eclipsing binaries within the TESS apertures identified in future \textit{Gaia} releases are mimicking a planetary transit. In particular, the 5$\%$ threshold ensures that even if a contaminant star lies within the TESS aperture (and is not accounted for by the TESS pipeline), the transit depth (and so the planet radius derived) will not be critically affected. After all previous filtering process, 26 intermediate-mass stars were added. Four of the previous were finally discarded because of the close companions ($\lesssim$ 1") detected in high-angular resolution imaging \citep{Lillo-Box24}. This leaves 22 additional intermediate-mass stars, almost doubling the initial sample identified above, and making a total of 47 sources included in the final sample.

In addition, we selected the 298 low-mass stars with CP or KP status in the TOIs catalog and inferred planetary sizes compatible with giant, gas made planets similar to those around the intermediate-mass sample. This list results from the step iv described above, constituting all TOIs with \textit{Gaia} DR3 masses $\leq$ 1.5 M$_{\odot}$ and confirmed short-period planets, in accordance with the Encyclopaedia of Exoplanetary Systems.

\section{Properties of the TOIs in this work}
\label{appendix_properties_sample}
Table \ref{table:int_mass_TOIS} lists main parameters of the 47 TOIs with intermediate-mass stars analysed in this work. Values for the stellar luminosities, temperatures, masses and radii were taken from the \textit{Gaia} DR3 catalog "I/355/paramp: 1D astrophysical parameters produced by the Apsis processing chain developed in \textit{Gaia} DPAC CU8". Among the different estimates for the effective temperatures available in that catalog, we selected the ones derived from BP/RP spectra. The main reason for this selection is that such temperature estimates cover a larger sample of stars than the rest. In addition, we checked that relative errors are mostly $<$ 10$\%$ when compared with effective temperatures derived from higher resolution spectra also available in that catalog. Errorbars for the stellar parameters refer to the lower (16$\%$) and upper (84$\%$) confidence levels listed in the \textit{Gaia} DR3 catalog. The planet classification status come from the "EXOFOP disposition" column in the TOIs catalog, from which the orbital periods and errors were also taken. Planet radii were derived from the R$_p$/R$_*$ values in the TOIs catalog and the stellar radii in the \textit{Gaia} DR3 catalog. The propagation of the corresponding uncertainties served to provide the final errorbars listed for the planet radii. Finally, orbital radii were derived from the Kepler's third law assuming that the planet mass is negligible compared with the stellar mass. Errorbars come from the propagation of the uncertainties in the orbital periods from the TOIs catalog and in the stellar masses from the \textit{Gaia} DR3 catalog.  

\begin{figure}
   \centering
   \includegraphics[width=9cm]{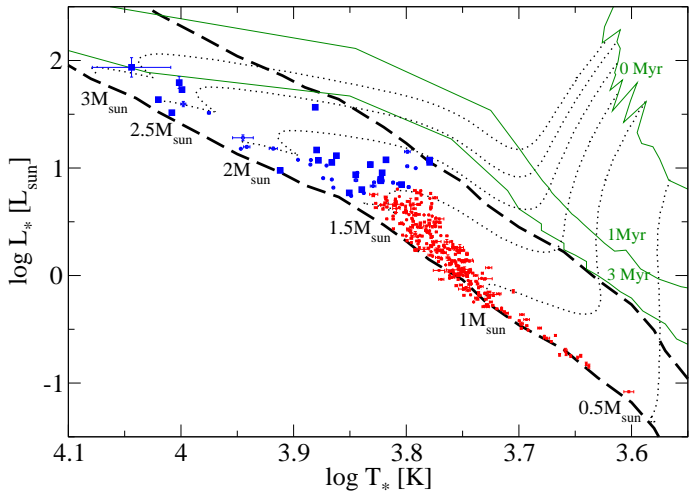}
      \caption{Stellar luminosity vs temperature for all TOIs in this work. Intermediate- and low-mass stars are in blue and red, respectively. Intermediate-mass stars hosting planets with candidate status in Table \ref{table:int_mass_TOIS} are indicated with squares. The evolution along the MS is bracketed by the dashed lines, the bottom line being the ZAMS and the top line the end of the MS phase. Representative pre-MS tracks (dotted lines with the corresponding stellar masses indicated) and isochrones at 0, 1 and 3 Myr (green solid lines) are also indicated.}
      \vspace{0.85cm}
         \label{fig:HR_diagram}
   \end{figure}

Figure \ref{fig:HR_diagram} shows the Hertzsprung-Russell (HR) diagram of the TOIs, adding to the intermediate-mass sample the 298 low-mass stars. All these host planets with CP and KP classification status (Appendix \ref{appendix_sample_selection}) and are plotted in red. Planets without such a classification around intermediate-mass stars (22 out of 47) are represented with blue squares. We also plotted the lower and upper dashed lines representing the beginning and the end of the MS phase \citep[for details about the MS evolution, see, e.g.,][]{Salaris06}. In addition, pre-MS tracks and isochrones from \citet{Siess00} are indicated for a representative set of stellar masses and ages (see Appendix \ref{appendix_pre-MS}).

\begin{figure}
   \centering
   \includegraphics[width=9cm]{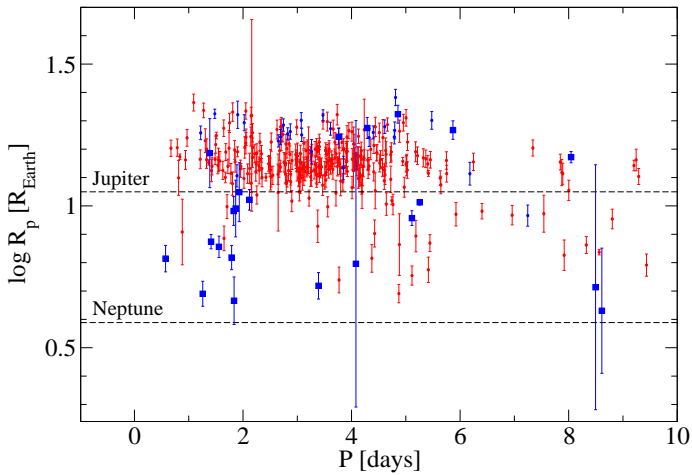}
      \caption{Planetary radii vs orbital periods for intermediate- (blue) and low-mass (red) TOIs in the sample. Planets around intermediate-mass stars with candidate status in Table \ref{table:int_mass_TOIS} are indicated with squares. For reference, the horizontal lines show the Neptune and Jupiter radii.} 
         \label{fig:planets_rad_period}
   \end{figure}

\begin{figure}
   \centering
   \includegraphics[width=9cm]{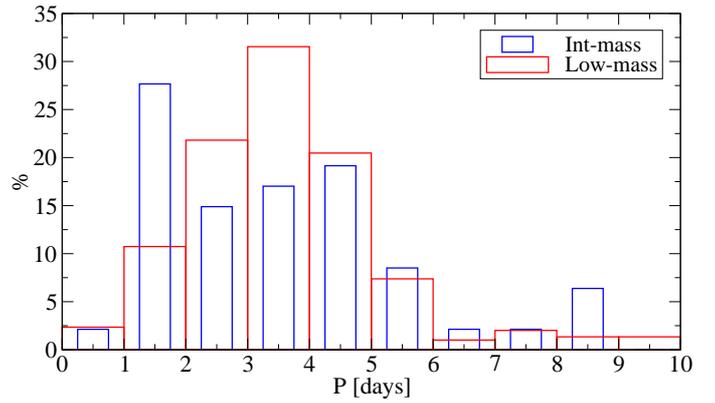}
      \caption{Distributions of planetary orbital periods. The intermediate- and low-mass samples are in blue and red, as indicated in the legend.} 
         \label{fig:planets_periods}
   \end{figure}

With respect to the properties of the planets, Fig. \ref{fig:planets_rad_period} shows the planetary radii and orbital periods for the TOIs in the sample. Planets around intermediate-and low-mass stars, as well as their classification status, are again indicated. Planet sizes in between that of Neptune and a few Jupiter radii homogeneously distribute along the whole range of periods for the intermediate- and low-mass samples. Figure \ref{fig:planets_periods} compares the distribution of orbital periods around intermediate- and low-mass stars. Planets around intermediate-mass stars peak at 1-2 days, contrasting with the “three-day pileup” of the population of hot Jupiters around low-mass stars observed here and in the literature \citep[e.g.,][and references therein]{Yee23}. Apart from this, the distributions of periods are similar for both stellar mass regimes. Indeed, a two-sample K-S test does not reject the null hypothesis that the orbital periods around intermediate- and low-mass stars are drawn from the same parent distribution, at a significance level given by p-value = 0.0957.

The similarity between the distributions of TESS planet sizes and periods explored in this work (Figs. \ref{fig:planets_rad_period} and \ref{fig:planets_periods}) suggests that the intermediate- and low-mass stars samples are similarly affected by potential observational biases. Thus, such biases should not originate the differences between both samples reported here (see Sect. \ref{Sect:results}).

\newpage
\onecolumn
\begin{table}
\centering
\caption{Sample of intermediate-mass stars}
\label{table:int_mass_TOIS}
\renewcommand\tabcolsep{2.2pt}
\renewcommand{\arraystretch}{1.05}
\begin{tabular}{|l|l|c|l|c|c|c|l|l|l|}
\hline
\hline
TOI&\textit{Gaia} DR3&log L$_*$&T$_*$&M$_*$&R$_*$&Class&P&R$_p$&r\\
%\hline
$\#$&$\#$&[L$_{\odot}$]&[K]&[M$_{\odot}$]&[R$_{\odot}$]&...&[10$^{-6}$ days]&[R$_{Earth}$]&[10$^{-4}$ au]\\
\hline
  159  & 4626444453769772160 & 0.769 $\pm$ 0.008 & 7089 $\pm$ 11 & 1.49 $\pm$ 0.04 & 1.61 $\pm$ 0.03 & PC& 3762837 $\pm$ 3.0 & 17.5 $\pm$ 1.2& 540 $\pm$ 5\\
  386  & 5497685201192526976 & 1.730 $\pm$ 0.016 & 9979 $\pm$ 21 & 2.51 $\pm$ 0.04 & 2.45 $\pm$ 0.05 & PC& 5111979 $\pm$ 5.0 & 9.1  $\pm$ 0.5& 790 $\pm$ 4\\
  508  & 3142847477107193344 & 0.906 $\pm$ 0.010 & 7479 $\pm$ 13 & 1.61 $\pm$ 0.04 & 1.69 $\pm$ 0.04 & KP& 4611733 $\pm$ 2.0 & 19.0 $\pm$ 1.0& 635 $\pm$ 5\\
  587  & 5702441891516232064 & 1.636 $\pm$ 0.011 & 10474$\pm$ 25 & 2.48 $\pm$ 0.04 & 2.00 $\pm$ 0.04 & PC& 8044239 $\pm$ 85  & 14.9 $\pm$ 0.7& 1064 $\pm$ 6\\
  624  & 3291455819447952768 & 1.281 $\pm$ 0.054 & 8811 $\pm$ 186& 2.00 $\pm$ 0.06 & 1.87 $\pm$ 0.02 & CP& 2744321 $\pm$ 2.0 & 19.3 $\pm$ 1.0& 483 $\pm$ 5\\
  625  & 3080104185367102592 & 1.032 $\pm$ 0.010 & 7675 $\pm$ 10 & 1.71 $\pm$ 0.04 & 1.85 $\pm$ 0.04 & CP& 4786958 $\pm$ 6.0 & 17.5 $\pm$ 1.0& 664 $\pm$ 5\\
  626  & 5617241426979996800 & 1.513 $\pm$ 0.014 & 9445 $\pm$ 29 & 2.25 $\pm$ 0.04 & 2.13 $\pm$ 0.05 & CP& 4401047 $\pm$ 1.0 & 18.2 $\pm$ 1.0& 689 $\pm$ 4\\
  748  & 6154982877300947840 & 0.887 $\pm$ 0.021 & 6993 $\pm$ 25 & 1.55 $\pm$ 0.04 & 1.89 $\pm$ 0.04 & KP& 2021958 $\pm$ 2.0 & 19.6 $\pm$ 1.3& 362 $\pm$ 3\\
  952  & 3188658560355725696 & 1.052 $\pm$ 0.028 & 7346 $\pm$ 14 & 1.70 $\pm$ 0.04 & 2.07 $\pm$ 0.05 & PC& 1790110 $\pm$ 10  & 6.6  $\pm$ 0.6& 344 $\pm$ 3\\
  1001 & 3067555424799970560 & 0.938 $\pm$ 0.010 & 6997 $\pm$ 17 & 1.59 $\pm$ 0.04 & 2.00 $\pm$ 0.04 & PC& 1931646 $\pm$ 5.0 & 11.2 $\pm$ 2.7& 354 $\pm$ 3\\
  1150 & 2064327278651198336 & 1.595 $\pm$ 0.050 & 9951 $\pm$ 50 & 2.32 $\pm$ 0.06 & 2.38 $\pm$ 0.02 & KP& 1481090 $\pm$ 0.0 & 21.1 $\pm$ 0.8& 337 $\pm$ 3\\
  1151 & 2033123654092592384 & 1.179 $\pm$ 0.005 & 8849 $\pm$ 3  & 1.92 $\pm$ 0.04 & 1.65 $\pm$ 0.03 & KP& 3474101 $\pm$ 0.0 & 20.9 $\pm$ 0.9& 558 $\pm$ 4\\
  1198 & 2054523105274216704 & 1.197 $\pm$ 0.019 & 8743 $\pm$ 57 & 1.93 $\pm$ 0.04 & 1.73 $\pm$ 0.04 & KP& 3612773 $\pm$ 3.0 & 18.7 $\pm$ 1.0& 573 $\pm$ 4\\
  1300 & 1612165353793791488 & 0.832 $\pm$ 0.012 & 6753 $\pm$ 39 & 1.50 $\pm$ 0.04 & 1.90 $\pm$ 0.04 & KP& 2871697 $\pm$ 1.0 & 18.3 $\pm$ 1.1& 452 $\pm$ 4\\
  1417 & 1905671319879758464 & 0.740 $\pm$ 0.007 & 7062 $\pm$ 11 & 1.47 $\pm$ 0.04 & 1.57 $\pm$ 0.03 & CP& 3072643 $\pm$ 1.0 & 20.0 $\pm$ 1.2& 470 $\pm$ 4\\
  1431 & 2188906825165621120 & 1.023 $\pm$ 0.005 & 7426 $\pm$ 1  & 1.68 $\pm$ 0.04 & 1.96 $\pm$ 0.04 & CP& 2650232 $\pm$ 0.0 & 16.9 $\pm$ 0.8& 446 $\pm$ 4\\
  1475 & 1940865385714676864 & 1.514 $\pm$ 0.010 & 10189$\pm$ 16 & 2.34 $\pm$ 0.04 & 1.84 $\pm$ 0.04 & PC& 8495351 $\pm$ 1270& 5.2  $\pm$ 5.1& 1082 $\pm$ 6\\
  1518 & 2210596444367488384 & 1.074 $\pm$ 0.006 & 7669 $\pm$ 9  & 1.74 $\pm$ 0.04 & 1.95 $\pm$ 0.04 & CP& 1902606 $\pm$ 7.0 & 21.0 $\pm$ 2.3& 361 $\pm$ 3\\
  1580 & 359678187913760384  & 0.853 $\pm$ 0.010 & 6411 $\pm$ 25 & 1.49 $\pm$ 0.04 & 2.17 $\pm$ 0.04 & KP& 4794773 $\pm$ 19  & 18.4 $\pm$ 1.2& 636 $\pm$ 6\\
  1599 & 328636019723252096  & 0.818 $\pm$ 0.008 & 7323 $\pm$ 10 & 1.54 $\pm$ 0.04 & 1.59 $\pm$ 0.03 & KP& 1219838 $\pm$ 17  & 18.1 $\pm$ 1.0& 258 $\pm$ 2\\
  1641 & 337169505562029568  & 1.936 $\pm$ 0.181 & 11064$\pm$ 887& 2.86 $\pm$ 0.22 & 2.59 $\pm$ 0.15 & PC& 1864673 $\pm$ 3.0 & 9.8  $\pm$ 2.2& 421 $\pm$ 11\\
  1904 & 6144125887169672064 & 0.823 $\pm$ 0.020 & 6279 $\pm$ 16 & 1.47 $\pm$ 0.04 & 2.18 $\pm$ 0.05 & KP& 4324771 $\pm$ 7.0 & 17.4 $\pm$ 1.3& 590 $\pm$ 5\\
  1907 & 6086712585429729536 & 0.955 $\pm$ 0.020 & 6973 $\pm$ 12 & 1.60 $\pm$ 0.04 & 2.06 $\pm$ 0.05 & KP& 5477443 $\pm$ 4.0 & 20.0 $\pm$ 1.4& 712 $\pm$ 6\\
  1916 & 5605119586158973440 & 1.181 $\pm$ 0.024 & 8281 $\pm$ 69 & 1.88 $\pm$ 0.04 & 1.86 $\pm$ 0.04 & KP& 1888199 $\pm$ 92  & 12.8 $\pm$ 1.2& 369 $\pm$ 3\\
  1924 & 5245968236116294016 & 1.077 $\pm$ 0.005 & 7872 $\pm$ 4  & 1.75 $\pm$ 0.04 & 1.86 $\pm$ 0.04 & KP& 2824069 $\pm$ 1.0 & 18.0 $\pm$ 0.9& 472 $\pm$ 4\\
  1933 & 2894378838731535872 & 0.865 $\pm$ 0.022 & 6482 $\pm$ 12 & 1.51 $\pm$ 0.04 & 2.15 $\pm$ 0.05 & KP& 3263316 $\pm$ 827 & 15.5 $\pm$ 1.5& 494 $\pm$ 4\\
  2014 & 1358614983131339392 & 1.017 $\pm$ 0.008 & 6296 $\pm$ 16 & 1.63 $\pm$ 0.04 & 2.71 $\pm$ 0.06 & KP& 4810114 $\pm$ 2.0 & 24.1 $\pm$ 1.6& 656 $\pm$ 5\\
  2036 & 1220999008291524224 & 0.846 $\pm$ 0.007 & 6370 $\pm$ 10 & 1.49 $\pm$ 0.04 & 2.17 $\pm$ 0.04 & PC& 1258333 $\pm$ 393 & 4.9  $\pm$ 0.5& 260 $\pm$ 2\\
  2390 & 4905056375217435904 & 0.881 $\pm$ 0.022 & 6647 $\pm$ 46 & 1.53 $\pm$ 0.04 & 2.08 $\pm$ 0.05 & PC& 2119457 $\pm$ 159 & 10.5 $\pm$ 0.9& 372 $\pm$ 3\\
  2401 & 5658673155407726464 & 1.074 $\pm$ 0.008 & 6016 $\pm$ 3  & 1.65 $\pm$ 0.04 & 3.17 $\pm$ 0.06 & PC& 5868198 $\pm$ 7.0 & 18.5 $\pm$ 1.4& 753 $\pm$ 6\\
  2432 & 16870631694208      & 1.052 $\pm$ 0.013 & 6014 $\pm$ 9  & 1.63 $\pm$ 0.04 & 3.09 $\pm$ 0.07 & PC& 1827380 $\pm$ 1140& 9.6  $\pm$ 1.1& 344 $\pm$ 3\\
  2461 & 3393939030531019520 & 0.767 $\pm$ 0.014 & 6881 $\pm$ 10 & 1.46 $\pm$ 0.04 & 1.70 $\pm$ 0.04 & KP& 6180268 $\pm$ 2.0 & 13.0 $\pm$ 1.2& 748 $\pm$ 7\\
  2541 & 2956998401052982016 & 1.565 $\pm$ 0.034 & 7600 $\pm$ 0  & 2.20 $\pm$ 0.05 & 3.23 $\pm$ 0.04 & PC& 1411842 $\pm$ 6.0 & 7.5  $\pm$ 0.4& 320 $\pm$ 2\\
  3678 & 116922311314227840  & 1.114 $\pm$ 0.017 & 7277 $\pm$ 13 & 1.75 $\pm$ 0.04 & 2.27 $\pm$ 0.05 & PC& 4854215 $\pm$ 58  & 21.1 $\pm$ 1.5& 675 $\pm$ 5\\
  4174 & 2260499600658158080 & 0.893 $\pm$ 0.022 & 6649 $\pm$ 7  & 1.54 $\pm$ 0.04 & 2.11 $\pm$ 0.05 & PC& 1556952 $\pm$ 6.0 & 7.2  $\pm$ 0.6& 303 $\pm$ 3\\
  4303 & 4814539625920138496 & 0.799 $\pm$ 0.006 & 6911 $\pm$ 9  & 1.49 $\pm$ 0.04 & 1.75 $\pm$ 0.04 & PC& 8611084 $\pm$ 37  & 4.3  $\pm$ 2.2& 938 $\pm$ 8\\
  4480 & 2052194374009877376 & 1.152 $\pm$ 0.025 & 6295 $\pm$ 36 & 1.74 $\pm$ 0.04 & 3.16 $\pm$ 0.08 & KP& 3849385 $\pm$ 32  & 13.6 $\pm$ 1.3& 578 $\pm$ 5\\
  4486 & 1835842749465226368 & 0.866 $\pm$ 0.009 & 6739 $\pm$ 14 & 1.52 $\pm$ 0.04 & 1.99 $\pm$ 0.04 & KP& 2691561 $\pm$ 2.0 & 17.5 $\pm$ 1.1& 435 $\pm$ 4\\
  4603 & 3402980516507429888 & 0.999 $\pm$ 0.007 & 6189 $\pm$ 4  & 1.61 $\pm$ 0.04 & 2.75 $\pm$ 0.06 & CP& 7245496 $\pm$ 275 & 9.2  $\pm$ 0.8& 858 $\pm$ 7\\
  4620 & 172870170218265088  & 1.072 $\pm$ 0.008 & 7551 $\pm$ 8  & 1.73 $\pm$ 0.04 & 2.01 $\pm$ 0.04 & PC& 1388749 $\pm$ 5.0 & 15.4 $\pm$ 4.3& 292 $\pm$ 2\\
  4987 & 3751877374435102720 & 1.032 $\pm$ 0.048 & 6791 $\pm$ 28 & 1.66 $\pm$ 0.05 & 2.37 $\pm$ 0.08 & PC& 4288970 $\pm$ 7.0 & 18.8 $\pm$ 1.6& 611 $\pm$ 6\\
  5074 & 651935220461650560  & 0.893 $\pm$ 0.008 & 7346 $\pm$ 3  & 1.59 $\pm$ 0.04 & 1.73 $\pm$ 0.04 & KP& 3080186 $\pm$ 10  & 17.8 $\pm$ 1.0& 483 $\pm$ 4\\
  5107 & 3438261890436625408 & 1.794 $\pm$ 0.112 & 10036$\pm$ 23 & 2.59 $\pm$ 0.08 & 2.61 $\pm$ 0.18 & PC& 1835437 $\pm$ 161 & 4.6  $\pm$ 0.9& 403 $\pm$ 4\\
  5381 & 926661154281525888  & 1.076 $\pm$ 0.014 & 6577 $\pm$ 18 & 1.69 $\pm$ 0.04 & 2.66 $\pm$ 0.06 & PC& 4080286 $\pm$ 34  & 6.2  $\pm$ 7.3& 595 $\pm$ 5\\
  5603 & 1574955715646199552 & 0.956 $\pm$ 0.026 & 6628 $\pm$ 6  & 1.59 $\pm$ 0.04 & 2.28 $\pm$ 0.06 & PC& 0573618 $\pm$ 2.0 & 6.5  $\pm$ 0.7& 158 $\pm$ 1\\
  6060 & 251128219566278144  & 1.168 $\pm$ 0.012 & 7579 $\pm$ 17 & 1.80 $\pm$ 0.04 & 2.23 $\pm$ 0.05 & PC& 3390631 $\pm$ 26  & 5.2  $\pm$ 0.6& 538 $\pm$ 4\\
  6291 & 216654200704824064  & 0.977 $\pm$ 0.019 & 8163 $\pm$ 18 & 1.72 $\pm$ 0.04 & 1.54 $\pm$ 0.04 & PC& 5255642 $\pm$ 39  & 10.3 $\pm$ 0.2& 709 $\pm$ 6\\

\hline
\hline
\end{tabular}
\begin{minipage}{18cm}

\textbf{Notes.} Columns 1 and 2 list the TOI and \textit{Gaia} DR3 identifications. Columns 3 to 6 list the stellar luminosity, temperature, mass and radius based on the \textit{Gaia} DR3 catalog of astrophysical parameters. Columns 7 and 8 list the planet classification status ("confirmed planet," "Kepler planet," or " planet candidate") and planet orbital period based on the TOIs catalog. The planet radii in col. 9 are derived from the R$_p$/R$_*$ ratios in the TOIs catalog and the stellar radii in col. 6. Planet orbital radii in col. 10 are derived from the Kepler`s third law, using stellar masses and periods in cols. 5 and 8.  
\end{minipage}
\end{table}   

\newpage
\twocolumn

\section{Pre-MS estimates}
\label{appendix_pre-MS}

\begin{figure}
   \centering
   \includegraphics[width=9cm]{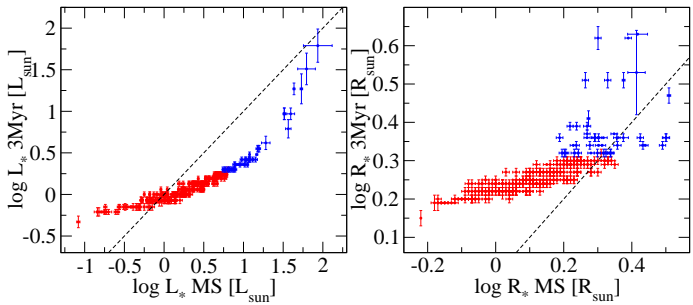}
      \caption{Stellar luminosities (left) and radii (right) during the MS are compared with those inferred at 3 Myr for the TOIs considered in this work. The dashed lines indicates equal values in both axes. Intermediate- and low-mass stars are in blue and red, respectively} 
         \label{fig:MS_vs_PMS}
   \end{figure}

The \citet{Siess00} isochrones provide, for a given age and metallicity, the corresponding values of the stellar parameters L$_*$, R$_*$, T$_*$, and M$_*$. In turn, the pre-MS evolution in the HR diagram can be inferred, for a given value of M$_*$ and metallicity, from the \citet{Siess00} evolutionary tracks. Based on the \textit{Gaia} DR3 stellar masses, the pre-MS values of L$_*$ and R$_*$ were inferred for each star in our sample using the 3 Myr isochrone from \citet{Siess00}, assuming solar metallicity. This isochrone is plotted in Fig. \ref{fig:HR_diagram}, along with several representative evolutionary tracks. From this graphical perspective, the pre-MS values for a given stellar mass are inferred from the point where the 3 Myr isochrone and the corresponding evolutionary track coincide. Figure \ref{fig:MS_vs_PMS} compares the MS luminosities and radii from \textit{Gaia} DR3 with those estimated during the pre-MS at 3 Myr. Errorbars in the y-axes reflect the uncertainties in the \textit{Gaia} DR3 stellar masses used to infer the pre-MS values of L$_*$ and R$_*$ based on the isochrone, and also take into account the fact that bins of 0.1M$_{\odot}$ were assumed to be interpolated. 

To estimate how disk dissipation timescales different than the typical 3 Myr affect our results and conclusions, we considered two cases.
First, it was assumed that disk dissipation in intermediate-mass stars is faster than in low-mass stars. This difference has been suggested in earlier works indicating that disks around intermediate- and low-mass stars dissipate mostly at $\sim$ 1 Myr and $\sim$ 3 Myr, respectively \citep[e.g.,][]{Ribas15}. The 1 Myr isochorne from \citet{Siess00} is plotted in Figure \ref{fig:HR_diagram} along with the 3 Myr isochrone. For a typical intermediate-mass star with M$_*$ = 2M$_{\odot}$, L$_*$ at 1 Myr is $\sim$ 0.2 dex larger than at 3 Myr. Because T$_*$ is slightly smaller at 1 Myr, R$_*$ at this age is larger by a factor $\sim$ 1.4 (0.1 dex) compared with that at 3 Myr. The net result is a slight displacement of the intermediate-mass stars to the right of both panels in Fig. \ref{fig:orbradius_gas_dust_preMS}, and to the left in Fig. \ref{fig:histogram_preMS}. In other words, our conclusion that planetary orbits around intermediate-mass stars tend to be smaller than the dust-destruction radius and consistent with small magnetospheres would be reinforced under this scenario.

\begin{figure}
   \centering
   \includegraphics[width=9cm]{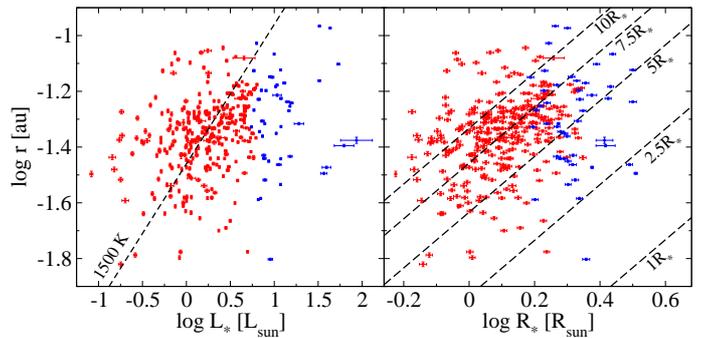}
      \caption{Planetary orbital radii versus MS stellar luminosities (left) and radii (right). Intermediate- and low-mass stars are in blue and red, respectively. In the left panel, the dashed line indicates the inner dust disk for a dust sublimation temperature of 1500 K. In the right panel, the dashed lines indicate the magnetospheric inner gas disk at 10, 7.5, 5, 2.5, and 1R$_*$. }
      \vspace{0.85 cm}
         \label{fig:orbradius_gas_dust_MS}
   \end{figure} 

\begin{figure}
   \centering
   \includegraphics[width=9cm]{Figures/histogram_seps_MS.eps}
      \caption{Distributions of planetary orbital radii in terms of of MS stellar radii. The intermediate- and low-mass samples are in blue and red, as indicated in the legend.} 
         \label{fig:histogram_MS}
   \end{figure}  

Second, it is explored the case in which the stars dissipate their disks at very late stages, as the stellar parameters become more similar to the final ones when they enter the MS. In particular, we consider the most extreme scenario by assuming that disk dissipation occurs when the stellar parameters are equal to the current, \textit{Gaia} DR3 ones. Figures \ref{fig:orbradius_gas_dust_MS} and \ref{fig:histogram_MS} are the MS versions of Figs. \ref{fig:orbradius_gas_dust_preMS} and \ref{fig:histogram_preMS}, where the \textit{Gaia} DR3 luminosities and radii have been used. According with Fig. \ref{fig:orbradius_gas_dust_MS}, all but one intermediate-mass stars host planets in orbits closer to the central source than the dust barrier, and the location of nearly $\sim$ 50$\%$ of the planets is consistent with small magnetospheres of $<$ 5R$_*$. In turn, around half of the low-mass stars host planets in orbits larger than the dust-destruction radius and the location of $\sim$ 90$\%$ of the planets are consistent with large magnetospheres $>$ 5R$_*$. The difference between the distributions of planetary orbital radii around intermediate- and low-mass stars in Fig. \ref{fig:histogram_MS} is even more pronounced than in Fig. \ref{fig:histogram_preMS}, with the planets around intermediate-mass stars dominating at orbits $<$ 6R$_*$ and the ones around low-mass stars dominating at larger distances. Indeed, the K-S test rejects that the samples in Fig. \ref{fig:histogram_MS} are drawn from the same parent distribution with a p-value of only 8.2 $\times$ 10$^{-8}$. Thus, the assumption that disks dissipate later than the typical 3 Myr does not essentially affect our results and conclusions.

In addition, compared with the previously discussed changes of the disk dissipation timescale, the use of metallicities that are different than solar or evolutionary tracks and isochrones different to those in \citet{Siess00} have a negligible effect on the pre-MS values inferred for L$_*$ and R$_*$ \citep[see, e.g.,][]{Siess00,Stassun14}. 

In summary, although different assumptions to infer the stellar luminosities and radii during disk dissipation have an effect on their specific values and related statistics, the general results and conclusions of this work remain unaltered.

\end{appendix}

\end{document}